\documentclass[preprint,12pt]{elsarticle}
\usepackage{url}
\usepackage{amssymb}
\usepackage{amsmath}
\usepackage{mathtools}

\usepackage{layout}
\usepackage{bm}
\usepackage{color}
\usepackage{float}
\usepackage{ulem}
\usepackage{siunitx}
\usepackage{cancel}
\newcommand{\A}[1]{{\textcolor{blue}{#1}}}

\newcommand{\D}[1]{{\textcolor{black}{#1}}}

\newcommand{\R}[1]{{\textcolor{black}{#1}}}

\newcommand{\RR}[1]{{\textcolor{black}{#1}}}

\journal{ }

\begin{document}
\begin{frontmatter}
\title{Experimental study on the relationship between extensional and shear rheology of \D{low-viscosity} power-law fluids}

\author{Yuzuki Matsumoto}
\author{Misa Kawaguchi}
\author{Yoshiyuki Tagawa}

\address{Department of Mechanical Systems Engineering, Tokyo University of Agriculture and Technology, 2-24-16 Nakacho, Koganei, 1848588 Tokyo, Japan}

\ead{tagawayo@cc.tuat.ac.jp}

\begin{abstract}
This paper investigates the relationship between extensional and shear viscosity of low-viscosity power-law fluids. \D{We \RR{show} the first experimental evidence of the conditions satisfying the same power exponents for extensional and shear viscosity, as indicated by the Carreau model.} The extensional and shear \D{viscosity} are respectively measured by capillary breakup extensional rheometry dripping-onto-substrate (CaBER-DoS) and by a shear rheometer for various Ohnesorge number $Oh$. \D{The viscosity ranges measured \RR{are} about $O(10^0)$ to $O(10^4)$ mPa$\cdot$s for shear viscosity and $O(10^1)$ to $O(10^3)$ mPa$\cdot$s for \R{apparent} extensional viscosity.}
Our experimental results show that, at least for the range of $Oh > 1$, the power-law expression for the liquid filament radius, \R{apparent} extensional viscosity, and shear viscosity holds, even for low-viscosity fluids \R{under our experimental condition\RR{s}}. 
\end{abstract}


\end{frontmatter}


\section{Introduction}
\label{S:1}

The rheology of complex fluids affects a wide range of phenomena, including on-demand printing technology~\cite{Onuki2018,Kamamoto2021}, inkjet printing~\cite{Sanjana2004}, spinning~\cite{Denn1980}, the investigation of food rheology~\cite{Berta2016}, and medical technology for patients with dysphagia~\cite{Hadde2019a,Hadde2019b}. 
In each of these cases, the evaluation of rheology, i.e. shear rheology and extensional rheology, is important. 
Shear rheology measurements can be performed on fluids with a wide range of shear viscosity. \D{For} various processes \D{such as printing technology}, it is important to evaluate the extensional rheology \D{of a wide range of fluids}, especially uniaxial extensional viscosity, in addition to the shear rheology. 

In particular, there has been little investigation of the relationship between the extensional viscosity and the shear viscosity for low-viscosity liquids.
Huisman \textit{et al.}~\cite{Huisman2012} investigated the relationship between the power-law exponent of the liquid filament radius \RR{using capillary breakup extensional rheometry (CaBER)} and the \RR{shear stress measured using rheometer} of power-law fluids.
\R{Although Huisman \textit{et al.}~\cite{Huisman2012} have investigated the power-law relationship between the filament radius and \RR{shear stress}, the extensional viscosity has not been quantified. }

Ng \textit{et al.}~\cite{Henry2020} examined the extensional and shear rheology of graphene oxide solutions, which are a type of power-law fluid, by varying the volume concentration $\phi$ from 0.02--0.4 vol\%. \D{The viscosity ranges measured were about $O(10^0)$ to $O(10^3)$ mPa$\cdot$s for shear viscosity and $O(10^2)$ to $O(10^3)$ mPa$\cdot$s for extensional viscosity.}
The exponents for the measured extensional viscosity (which was measured using CaBER) and shear viscosity were in good agreement at high volume concentrations ($\phi>$ 0.15 vol\%).
Ng \textit{et al.}~\cite{Henry2020} stated that the extensional rate of low-concentration solutions is very fast and difficult to measure using CaBER, and proposed the use of CaBER dripping-onto-substrate (CaBER-DoS) to solve this problem. 

From the Carreau model~\cite{Carreau1979}, the power-law exponents for the filament radius, extensional viscosity, and shear viscosity were found to be closely related. However, the applicable conditions for the Carreau model are unclear due to the lack of extensional viscosity measurements in low-viscosity fluids.

Several extensional rheology measurement methods have been proposed in previous studies. 
CaBER ~\cite{Anna2001} and filament stretching extensional rheometry (FiSER)~\cite{mckinley2002} are popular methods~\cite{Sur2018} for measuring extensional rheology.
In CaBER and FiSER, the liquid fills a gap between two plates, and the plates are then pulled apart to extend the liquid.
FiSER is typically applied to polymeric solutions with relatively high shear viscosity and high extensional viscosity, because the minimum and maximum  strain rates applied to the liquid are fixed and FiSER can be used when the zero-shear viscosity $\eta_0>1$ Pa$\cdot$s.
CaBER can measure relaxation times as low as $\lambda=1$ ms for low-viscosity fluids (as low as 70 mPa$\cdot$s)~\cite{Vadillo2012}.
To measure liquids with viscosities and relaxation times below the limits of FiSER and CaBER, a Rayleigh--Ohnesorge jetting extensional rheometer (ROJER)~\cite{Keshavarz2015,mathues2018} can be used.
The ROJER generates a liquid jet, the extensional behavior of which is observed. 
\R{There are reports of ROJER measurements with relatively low zero-shear viscosity  $\eta_0 = 2.9$ mPa$\cdot$s \cite{greiciunas2017design} and relatively short relaxation time of $\lambda = 60$ $\mu$s \cite{Keshavarz2015}.} The limitations of existing measurement methods mean that few studies have evaluated the extensional rheology of low-viscosity solutions.

Recently, CaBER-DoS has been used to measure solutions with even lower viscosities and shorter relaxation times than permitted by the ROJER.  
While CaBER utilize the self-thinning of a stretched liquid bridge formed by applying a discrete step strain to a drop between two parallel plates, CaBER-DoS utilize the simple dripping system~\cite{dinic2017pinch}.
\RR{CaBER-DoS facilitates visualization of necking because the location of the thinning neck is fixed in the Eulerian location. }
Thus, CaBER-DoS achieves a higher temporal resolution and can accurately measure the liquid filament radius.
With CaBER-DoS, the zero-shear viscosity can be measured to a minimum of $\eta_0 \sim 1$ mPa$\cdot$s and the relaxation time reaches a minimum of $\lambda \sim O$({$\mu$}\rm{s})~\cite{Sur2018,Dinic2017}.
Nevertheless, extensional viscosity measurements have generally been performed on relatively high-viscosity liquids such as polymeric solutions. 
Therefore, it is necessary to conduct extensional rheology measurements on low-viscosity solutions and analyze their extensional rheological characteristics~\cite{Bhattacharjee2011}.

The purpose of this paper is to experimentally clarify the applicable conditions of the relationship between the extensional and shear rheology in power-law fluids. The extensional viscosity is measured using CaBER-DoS and the shear viscosity is measured using a shear rheometer.
The theoretical relationships between the extensional and shear viscosities given by the Carreau model~\cite{Carreau1979,Yildirim2001} are compared with the measured results to investigate the conditions under which the theory is applicable.
A previous numerical study~\cite{Suryo2006} suggested that the power-law relationship holds when the Ohnesorge number $Oh$ is greater than 1.
$Oh$ is the ratio of the viscous force to the product of the inertia and the surface tension, i.e., $Oh = \eta_0/\sqrt{\rho \gamma R_0}$, where $\eta_0$ is zero-strain-rate viscosity, $R_0$ is the nozzle radius, $\gamma$ is the surface tension, and $\rho$ is the density.
We experimentally clarify the range of $Oh$ under which the exponential relationship holds, including the shear viscosity as well as the extensional viscosity.
To investigate the rheology under different values of $Oh$, the viscosity of the solution is varied by changing its \R{dilution}.

The remainder of this paper is organized as follows.
In Section 2, the methodology of CaBER-DoS (\S 2.1) is explained and the extensional and shear viscosity (\S 2.2) are defined. In Section 3, the experimental setup and methodology are introduced. 
Section 4 presents and analyzes the experimental results. The power-law exponents are investigated for the filament radius (\S 4.1), extensional viscosity (\S 4.2), and shear viscosity (\S 4.3) for various $Oh$, and the applicable conditions are discussed (\S 4.4). Finally, Section 5 summarizes our findings. 

\section{Theory}
\label{S:2}
\subsection{CaBER-DoS}
In CaBER-DoS \cite{Dinic2017}, the temporal evolution of a representative length of a liquid filament, i.e., the filament radius $R$, is measured.
By considering the resistance force on the liquid filament against the surface tension, the equation for the time evolution of the filament radius $R$ is derived.

In the inertia-capillary regime, where the \RR{inertial acceleration} is dominant as a resisting force against the surface tension, the filament radius decays with time as follows \cite{mckinley2005visco}
\begin{eqnarray}
R/R_0 \propto \left( \frac{\gamma}{\rho {R_0}^3} \right)^{1/3} (t_c - t)^{2/3},\label{eq:inertia}
\end{eqnarray}
where $R_0$ is the nozzle radius and $t_c$ is the pinch-off time.
Equation~\eqref{eq:inertia} assumes that the resistance force against the surface tension exerted on the liquid filament is only the inertia\RR{l acceleration} from $t = 0$ to the pinch-off at $t = t_c$.

In the visco-capillary regime, where the viscous force is dominant as a resisting force against the surface tension, the filament radius decays linearly with time as follows \cite{mckinley2000extract}: 
\begin{eqnarray}
R/R_0 = \frac{2 \chi -1}{6} \frac{\gamma}{\eta_S R_0} (t_c - t).\label{eq:viscous}
\end{eqnarray}
Equation~\eqref{eq:viscous} assumes that the resistance against the surface tension exerted on the liquid filament is solely provided by the viscous force from $t = 0$ to the pinch-off at $t = t_c$. 
The parameter $\chi$ in Eq.~\eqref{eq:viscous} is expressed as
\begin{equation}
\chi(t) = \frac{F_z(t)}{2 \pi \gamma R(t)},
\label{eq:chi}
\end{equation}
where $F_z(t)$ is the unknown tensile force acting along the column.
When $\chi=1$, the liquid filament is perfectly cylindrical.
The tension force $F_z(t)$ varies with time and the filament shape.
As the filament approaches a cylindrical shape, $F_z(t)$ approaches the axial force $2 \pi \gamma R(t)$ caused by the surface tension:
\begin{equation}
\lim_{R \to 0} F_z(t) = 2 \pi \gamma R(t).
\end{equation}

\RR{For a power-law fluid, the decrease in local effective viscosity at the necking point leads to necking rate acceleration. T}he radius $R$ is expressed as follows \RR{\cite{mckinley2005visco,Clasen2012}}:
\begin{eqnarray}
R/R_0 \propto (t_c - t)^n, \label{eq11}
\end{eqnarray}
where $n$ is the power-law exponent.
From the CaBER-DoS experiments, the extensional rate $\dot{\varepsilon}$, Hencky strain $\varepsilon$, and \R{apparent} extensional viscosity $\eta _E$ can be calculated using the measured filament radius $R$~\cite{Sur2018,Dinic2017,mckinley2000extract}:
\begin{equation}
\dot{\varepsilon}
= -\frac{2}{R}\frac{dR}{dt},\ \ 
\varepsilon = 2 \ln{\frac{R_0}{R}},\ \ 
\eta_E = -\frac{\gamma}{2 (dR/dt)}.
\label{eq:Ex_rheology}
\end{equation}
The \R{apparent} extensional viscosity in Eq.~\eqref{eq:Ex_rheology} assumes a cylindrical filament ($\chi = 1$). 
In the case of power-law fluids, the actual liquid filament is not cylindrical, but the filament radius changes along the $z$-direction. 
The force balance in the liquid filament considering the curvature is as follows:
\begin{equation}
\frac{\gamma}{R} - \frac{ F_z(t)}{\pi R(t)^2} = \eta_E\dot{\varepsilon} .
\label{eq:balance_w_curvature}
\end{equation}

Using $\chi$ to account for the curvature, the \R{apparent} extensional viscosity can be expressed by substituting Eq.~\eqref{eq:chi} into Eq.~\eqref{eq:balance_w_curvature}:
\begin{equation}
\eta_E = \frac{(2 \chi -1)\gamma}{2 (dR/dt)}.
\label{eq:exvis}
\end{equation}

In \RR{a} previous \R{study}, Eggers \cite{Eggers1993} proposed that $\chi = 0.5912$.
The inertial \RR{acceleration} cannot be neglected because the filament rapidly becomes thinner just before pinch-off, and so the \R{strain rate} is large while the \R{apparent} extensional viscosity is small.
Therefore, numerical calculations were used to obtain a universal similarity solution that balances inertial, capillary, and viscous effects \cite{Eggers1993}.
According to Brenner \textit{et al.}~\cite{Brenner1996}, the solution of Eggers \cite{Eggers1993} is a special case because it is the least unstable to finite-amplitude perturbations, and $\chi$ has countably infinite similarity solutions. Papageorgiou~\cite{Papageorgiou1995} numerically obtained the value $\chi = 0.7127$ from the unsteady Stokes equation for the case in which the inertial \RR{acceleration} is neglected and only the viscous force is applied to the liquid.
This is supported by McKinley and Tripathi \cite{mckinley2000extract}, who stated that $\chi = 0.7127$ is the most appropriate value under typical experimental conditions.

Recently, $\chi=0.7127$ was adopted in CaBER experiments using graphene oxide, which is a power-law fluid~\cite{Henry2020}. In our experiments, we adopted the value of $\chi = 0.7127$ derived by Papageorgiou~\cite{Papageorgiou1995}, as used by McKinley and Tripathi~\cite{mckinley2000extract} and Ng \textit{et al.}~\cite{Henry2020}.

\subsection{Extensional viscosity and shear viscosity in three-parameter Carreau model}
In this subsection, the \RR{expressions for} both the shear and extensional viscosity are derived using the three-parameter Carreau model \cite{Carreau1979}.
The three-parameter Carreau model is an effective method for expressing the shear-thinning of fluids.
The Carreau model is one of the Generalised Newtonian Fluid (GNF) models. GNF models assume that the viscosity of non-Newtonian fluid is independent of deformation history and it is modeled inelastically as a function of the rate of strain.
As shown below, the GNF models are a function of the magnitude of the strain rate tensor, which is related to the second principal invariant of this tensor.

Yildirim \textit{et al.}~\cite{Yildirim2001} and Doshi \textit{et al.}~\cite{Doshi2003} have used this model to consider the viscosity in uniaxial extension.
In the three-parameter Carreau model, viscosity is expressed as 
\begin{eqnarray}
\eta=\R{\eta_0}(1-\beta)[1+(\alpha \hat{\dot{\gamma}})^2]^{(n-1)/2}+\beta \R{\eta_0}, 
\label{eqCarreaumodel}
\end{eqnarray}
where \R{${\eta}$ is the apparent viscosity, $\eta_0$ is the zero-strain-rate viscosity,} $\beta$\RR{$\eta_0$} is the viscosity in the limit as the deformation rate goes to infinity ($0<\beta\leq 1$), $\alpha$ is a constant with the dimensions of time, $\hat{\dot{\gamma}}$ is deformation rate, and $n$ is the power-law index.
A smaller characteristic shear rate $\alpha^{-1}$ makes it easier to maintain a high viscosity as the strain rate increases.

First, we derive the shear viscosity by considering that $\hat{\dot{\gamma}}^2$ in Eq.~\eqref{eqCarreaumodel} is expressed through the second invariant.~
In the Cartesian coordinate system, the velocity gradient tensor $\bf{\nabla} \bm{\upsilon}$ in shear flow is
\begin{eqnarray}
\bf{\nabla} \bm{\upsilon} 
= \left( 
\begin{array}{ccc}
0 & \displaystyle \frac{\partial \upsilon_x}{\partial y} & 0 \\
& \\
\displaystyle \frac{\partial \upsilon_y}{\partial x} & 0 & 0 \\
& \\
0 & 0 & 0
\end{array}
\right). \label{eq8}
\end{eqnarray}

The second invariant of the rate of deformation tensor ${\textbf{II}_S}$ in shear flow is

\begin{eqnarray}
\RR{\text{II}_S}
= \frac{1}{\RR{2}} \left( \frac{\partial \upsilon_x}{\partial y} + \frac{\partial \upsilon_y}{\partial x} \right)^2
= \frac{1}{\RR{2}} {\dot{\gamma}}^2, \label{eq18}
\end{eqnarray}
where $\dot{\gamma}$ is the strain rate. \RR{Note that the definition of the second invariant of the tensor is taken from the textbook \cite{bird1987dynamics} and their notation for $II$ is used.}

In GNF model \cite{bird, poole2023inelastic}, the stress tensor $\bm{\tau}$ is expressed as 

\begin{eqnarray}
\bm{\tau} = \eta (\bf{\nabla} \bf{\upsilon} + \bf{\nabla} \bf{\upsilon}^T) \equiv \eta \bm{\dot{\gamma}},
\label{GNFmodel}
\end{eqnarray}
where 
we introduced the strain rate tensor $\bm{\dot{\gamma}}$ = $(\bf{\nabla} \bf{\upsilon} + \bf{\nabla} \bf{\upsilon}^T)$.
The non-Newtonian viscosity $\eta$ is the function of shear rate, which in general can be written as the magnitude of the strain rate tensor $\dot{\gamma}$ = $\sqrt{\frac{1}{2} (\bm{\dot{\gamma}} \colon \bm{\dot{\gamma}})}$~\cite{bird}.
\R{The deformation rate $\hat{\dot{\gamma}}$ is expressed using the second invariant \RR{${\text{II}_S}$}, }
\begin{eqnarray}
\hat{\dot{\gamma}} = \R{\sqrt{4\left| \RR{{\text{II}_S}} \right|}} ~= \RR{\sqrt{2}}\R{\dot{\gamma}}.
~
\label{eqSecInS}
\end{eqnarray}

Thus, from Eqs.~\eqref{eqCarreaumodel},~\eqref{eq18}, and \eqref{eqSecInS}, the shear viscosity is obtained as follows:
\begin{eqnarray}
{\eta_S}\R{/\eta_0} = (1-\beta)[1+\RR{{2}} \alpha^2 {\dot{\gamma}}^2]^{(n-1)/2}+\beta.
\label{eq:shearvis}
\end{eqnarray}

Second, we derive the extensional viscosity by considering the liquid filament to be a thin axisymmetric liquid column in a cylindrical coordinate system. The velocity gradient tensor $\nabla \bm{\upsilon}$ is given by

\begin{eqnarray}
\nabla \bm{\upsilon} 
&=& 
\left(
\begin{array}{ccc}
\displaystyle \frac{\partial \upsilon_r}{\partial r} & \displaystyle \frac{1}{r} \frac{\partial \upsilon_r}{\partial \theta} - \frac{\upsilon_{\theta}}{r} & \displaystyle \frac{\partial \upsilon_r}{\partial z}\\
& \\
\displaystyle \frac{\partial \upsilon_{\theta}}{\partial r} & \displaystyle \frac{1}{r} \frac{\partial \upsilon_{\theta}}{\partial \theta} + \frac{\upsilon_r}{r} & \displaystyle \frac{\partial \upsilon_{\theta}}{\partial z}\\
& \\
\displaystyle \frac{\partial \upsilon_z}{\partial r} & \displaystyle \frac{1}{r} \frac{\partial \upsilon_z}{\partial \theta} & \displaystyle \frac{\partial \upsilon_z}{\partial z}
\end{array}
\right), \label{eq2}
\end{eqnarray}
where $\upsilon_r, \upsilon_{\theta}$, and $\upsilon_z$ are the velocity components in the $r,\theta$, and $z$ directions, respectively.
As the liquid column is symmetric about the $z$-axis, 
\begin{eqnarray}
\bf{\nabla} \bm{\upsilon} 
= \left( 
\begin{array}{ccc}
\displaystyle \frac{\partial \upsilon_r}{\partial r} & 0 & \displaystyle \frac{\partial \upsilon_r}{\partial z}\\
& \\
0 & \displaystyle \frac{\upsilon_r}{r} & 0\\
& \\
\displaystyle \frac{\partial \upsilon_z}{\partial r} & 0 & \displaystyle \frac{\partial \upsilon_z}
{\partial z} 
\end{array}
\right). \label{eq3}
\end{eqnarray} 
During extension, the radius is deformed in the negative direction. 
As $z=0$ at the nozzle end, the $z$-axis direction is also deformed in the negative direction. 
The second invariant of the deformation rate tensor \RR{${\text{II}_E}$} is given by
\begin{eqnarray}
\RR{\text{II}_E}
=  \RR{-2}\left( \frac{\upsilon_r}{r} \frac{\partial \upsilon_z}{\partial z} 
+ \frac{\upsilon_r}{r} \frac{\partial \upsilon_r}{\partial r} 
+ \frac{\partial \upsilon_r}{\partial r}\frac{\partial \upsilon_z}{\partial z} \right)
\RR{+ \frac{1}{2}}\left( \frac{\partial \upsilon_r}{\partial z} + \frac{\partial \upsilon_z}{\partial r} \right)^2.
\label{eq5} 
\end{eqnarray}
\RR{This is based on the definition denoted as $II$ in the book \cite{bird1987dynamics}.}

\R{Although it is assumed that the liquid column is perfectly cylindrical in this theory, we used the parameter $\chi$ = 0.7127 to consider the curvature of liquid filament in our experiment as described in \S 2.1.}

\begin{eqnarray}
\upsilon_z &=& z \dot{\varepsilon}
= z\frac{\partial \upsilon_z}{\partial z}
= -\frac{2z}{R}\frac{dR}{dt}, \nonumber \\
\upsilon_r &=& -\frac{r}{2} \dot{\varepsilon}
= -\frac{r}{2} \left( -\frac{2}{R}\frac{dR}{dt} \right)
= \frac{r}{R}\frac{dR}{dt} \nonumber ,
\end{eqnarray}
where the rate of strain in extensional deformation is defined as $\dot{\varepsilon}$ = $\partial \upsilon_z / \partial z$.
Thus, the second invariant of the rate of deformation tensor is given by
\begin{align}
\RR{\text{II}_E}
&
\begin{multlined}[t][7cm]
= -\RR{2}\left[ \frac{1}{r} \left( \frac{r}{R}\frac{dR}{dt} \right ) \left( -\frac{2}{R}\frac{dR}{dt} \right )+ \frac{1}{r} \left( \frac{r}{R}\frac{dR}{dt} \right )\left( \frac{1}{R}\frac{dR}{dt} \right ) 
+ \left( \frac{1}{R}\frac{dR}{dt} \right ) \left( -\frac{2}{R}\frac{dR}{dt} \right ) \right]
\end{multlined} \nonumber
\\
& =
\RR{6} \left( \RR{-\frac{1}{R}}\frac{dR}{dt} \right )^2 \nonumber \\
& =
\RR{\frac{3}{2}} \left( \RR{-\frac{2}{R}}\frac{dR}{dt} \right )^2 \nonumber \\
& = \RR{\frac{3}{2}} \dot{\varepsilon}^2.\label{eq24} 
\end{align}
As same as the shear flow, using the second invariant \RR{${\text{II}_E}$} in uniaxial extensional flow, the deformation rate $\hat{\dot{\gamma}}$ is expressed as
\begin{eqnarray}
\hat{\dot{\gamma}} = \sqrt{\R{4\left| \RR{{\text{II}_E}} \right|}}  = \R{\RR{\sqrt\frac{3}{2}}\dot{\varepsilon}}. \label{eqSecInE}
\end{eqnarray}
 
From Eqs.~\eqref{eqCarreaumodel},~\eqref{eq24}, and~\eqref{eqSecInE}, the extensional viscosity is given by
\begin{eqnarray}
{\eta_E}\R{/\eta_0} = (1-\beta)[1+\frac{3}{2} \alpha^2 
{\dot{\varepsilon}}^2]^{(n-1)/2}+\beta.
\label{eq:extensionalvis}
\end{eqnarray}

Suryo and Basaran~\cite{Suryo2006} suggested a transition from the potential flow regime to the non-slender viscous power-law regime for $n \leq 0.54$ when $Oh > 1$. From Eqs.~\eqref{eq:shearvis} and \eqref{eq:extensionalvis}, the extensional viscosity and shear viscosity have similar strain-rate dependence and the power exponents for the shear and extensional viscosity are equal.


\section{Experimental method}
\begin{figure}[H]
\centering
\includegraphics[width=\linewidth]{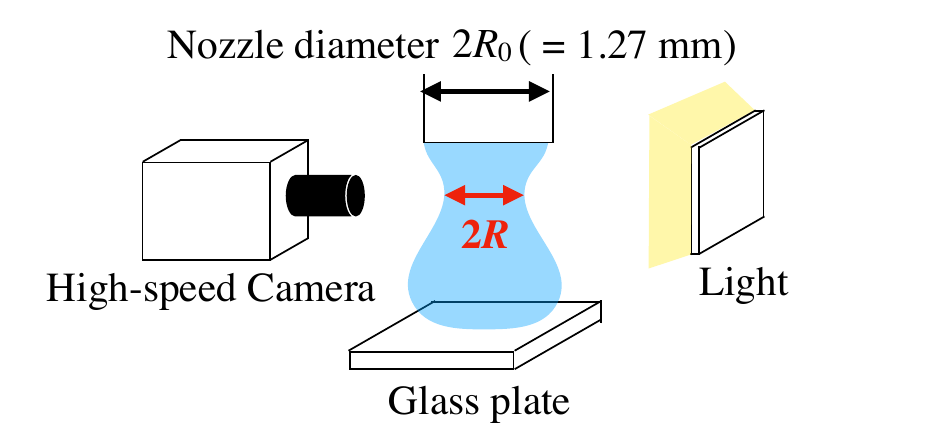}
\caption{Schematic view of CaBER-DoS setup. $R_0$ is the nozzle radius. $R$ is the smallest filament radius.}
\label{Fig:caber}
\end{figure}

A schematic of the CaBER-DoS setup used to measure the extensional rheology is shown in Fig.~\ref{Fig:caber}. 
This system simply consists of a nozzle and a glass substrate.
A syringe is connected to a syringe pump (Pump 11 Elite, Harvard Apparatus) and filled with a test liquid. 
The nozzle (PN-18G-B, Musashi Engineering, outer diameter $R_0$ = 635 $\mu$m) of the syringe is placed vertically.
A hydrophilic glass substrate is placed below the nozzle at a distance of $H$.
The optimal aspect ratio between the nozzle radius $R_0$ and distance $H$ is known to be around 6 (i.e., $H\sim 6R_0$)~\cite{Dinic2017}.
The droplet grows at the nozzle until the bottom of the droplet contacts the glass substrate.
The liquid then spreads on the glass substrate because of the wettability of the glass, and a liquid bridge forms between the nozzle and the substrate.
This filament thinning was captured by a high-speed camera (\R{Photron}, FASTCAM SA-X or SA-Z, 3.5 $\mu$m/pix, 500--10,000 fps).
The experiments were conducted at 20$^{\circ}$C.
The obtained images were analyzed using MATLAB to measure the filament radius $R$.
When the filament radius is equal to the nozzle radius ($R = R_0$), the time $t$ is defined as 0.

The measured liquid-filament radius $R$ was fitted using Eq.~\eqref{eq11}.
The fitting range is based on the power-law approximation of the liquid filament radius.
The early stages of necking are dominated by gravitational drainage~\cite{mckinley2000extract}. 
Gravitational effects can be neglected when the Bond number $Bo  ~= \rho g (2R)^2 / \sigma  \ll 1$.
If gravitational drainage is no longer dominant when $Bo \simeq 0.1$ and viscous forces become dominant, then the liquid filament radius can be governed by similarity solution of Papageorgiou~\cite{Papageorgiou1995} ($\chi$ = 0.7127) in the range 
\begin{equation}
2R \leq \sqrt{\frac{\sigma}{10 \rho g}}.
\label{eq:0023}
\end{equation}

Immediately before pinch-off, the extensional rate increases rapidly because the liquid filament becomes very thin, and the extensional viscosity decreases~\cite{Clasen2012}. 
Therefore, the force acting on the liquid filament changes from viscous to inertial.
The threshold of approximation just before pinch-off is determined by the confidence interval, which expresses the degree of uncertainty of the approximation.
In this experiment, the confidence interval was set to 95\%, indicating that there is a 95\% possibility that the new observation falls between the upper and lower bounds of the confidence interval. First, from Eq.~\eqref{eq:0023}, we approximated the range in which the confidence interval is considered to be narrow as follows:
\begin{equation}
\frac{1}{2R_0} \sqrt{\frac{\sigma}{10 \rho g}}-0.2 < \frac{R}{R_0} < \frac{1}{2R_0} \sqrt{\frac{\sigma}{10 \rho g}}.
\label{eq:0024}
\end{equation}
\R{Note that the 0.2 for lower limit is determined as a narrow enough range.} The approximation range was then gradually extended. 
The approximation range was set to be the widest \RR{without} exceeding the confidence \RR{bound} of 95\% \RR{for power-law index}. This procedure was conducted using Curve Fitting Toolbox\RR{,} an application software of MATLAB. The uncertainty for the exponent obtained using this method is approximately 6.5\% within the 95\% confidence interval \RR{of necking radius. }

Using the measured liquid filament radius, the extensional rate was calculated from Eq.~\eqref{eq:Ex_rheology} and the \R{apparent} extensional viscosity was calculated from Eq.~\eqref{eq:exvis}. 
The extensional rate and \R{apparent} extensional viscosity were calculated using the liquid filament radius within the power-law approximation.
\R{The exponent in Eq.~\eqref{eq:extensionalvis} was changed as }

\begin{eqnarray}
\R{l = n-1,}
\label{eq:l}
\end{eqnarray}
\R{and the measured apparent extensional viscosity was fitted by}
\begin{eqnarray}
{\eta_E}\RR{/\eta_{0,E}} = (1-\RR{\beta_E})[1+\frac{3}{2} \RR{\alpha_E}^2 
{\dot{\varepsilon}}^2]^{l/2}+\RR{\beta_E},
\label{eq:extensionalvis_fit}
\end{eqnarray}
and the value of the power exponent $l$ was obtained. \RR{Note that to distinguish the variables of $\eta_{0}$, $\alpha$ and $\beta$ from shear and extensional viscosity, $\eta_{0,E}$, $\alpha_E$ and $\beta_E$ are defined for extensional viscosity.}

The shear viscosity was measured using a shear rheometer (MCR302, Anton Paar) \R{equipped with cone plate (CP50-0.5, Anton Paar)with angle of 0.506 deg} over a shear rate range of 0.01--10,000 $\rm{s}^{-1}$ \R{at 20$^{\circ}$C}.

\R{There are \RR{detrimental effects} to the viscosity measurement using rheometer, such as mechanical limitations and secondary flows.}
When the torque of the shear rheometer is too small, the sensitivity of the measurement becomes worse and the measured shear viscosity becomes unreliable. 
The range of reliable shear viscosity measurements is given by~\cite{Gaillard2019}
\begin{equation}
\eta_S \RR{>} \frac{3 T_{min}}{2 \pi {R_{plate}}^3 \RR{\dot{\gamma}}},
\label{eq:23}
\end{equation}
where $T_{min}$ is the minimum torque and $R_{plate}$ \R{= 25 mm} is the \R{cone} plate radius of the shear rheometer.

\R{Secondary flow occurs at high shear rate. The secondary flow increase measured torque, therefore, increase the apparent viscosity incorrectly. The upper limit \cite{ewoldt2015experimental} can be estimated as}
\begin{equation}
\R{\eta_\mathrm{max} = \frac{\RR{\theta}^3 R_{plate}^2}{Re_\mathrm{crit}}\rho\dot{\gamma},}
\label{eq:vis_upper_limit} 
\end{equation}
\R{where the \RR{$\theta$} [rad] is the angle between cone and plate, $\rho$ is the density of the fluid, and $Re_\mathrm{crit}$ is the critical Reynolds number, which is identified by torque error. When $Re_\mathrm{crit}$ is approximated as 4 with torque error of 1\% \cite{ewoldt2015experimental}, the measured range of shear rate is not included region where secondary flow affects.}

\R{The exponent in Eq.~\eqref{eq:shearvis} was changed as }

\begin{eqnarray}
\R{m = n-1,}
\label{eq:m}
\end{eqnarray}
\R{and the measured shear viscosity was fitted by}
\begin{eqnarray}
{\eta_S}\RR{/\eta_{0,S}} = (1-\RR{\beta_S})[1+\RR{2} \RR{\alpha_S}^2 {\dot{\gamma}}^2]^{m/2}+\RR{\beta_S},
\label{eq:shearvis_fit}
\end{eqnarray}
and the value of the power exponent $m$ was obtained. \RR{Note that to distinguish the variables of $\eta_{0}$, $\alpha$ and $\beta$ from shear and extensional viscosity, $\eta_{0,S}$, $\alpha_S$ and $\beta_S$ are defined for shear viscosity.}

\R{The extensional and shear viscosity are similar strain-rate dependence as shown in Eqs.~\eqref{eq:extensionalvis_fit} and ~\eqref{eq:shearvis_fit}.} The power exponents for the shear viscosity and extensional viscosity are theoretically equal, so we expect
\begin{eqnarray}
m/l=\R{\frac{n-1}{n-1}=} 1.
\label{eq:relationship_m_l}
\end{eqnarray}
We regard \RR{Eqs.~\eqref{eq:l} and \eqref{eq:relationship_m_l}} to be the fundamental relations, and investigate the conditions under which these equations are valid.
For a Newtonian fluid in which \RR{inertial acceleration is dominant} on the liquid filament, we expect that $n=2/3$, $m=0$, and $l=0$, and so \RR{Eq.~\eqref{eq:l}} no longer holds.

To measure the surface tension of the solutions, the pendant drop method was used. 
This method determines the surface tension by acquiring images of pendant droplets and applying the Young--Laplace equation, which considers the equilibrium between the gravitational deformation of the droplet and the surface tension, to the droplet shape. 
The process of determining the surface tension using the Young--Laplace equation can be automated by using a high-quality digital camera and a computer.
In this study, the Opendrop software~\cite{Berry2015} was used to determine the surface tension. The surface tension values of the test liquids are listed in Table \ref{Table:surface_tension}. \R{Note that the measured data includes the error 1\%. The measured surface tension for water  at 20$^{\circ}$C as 74 mN/m and reference value is 72.88 mN/m at 293 K \cite{deen2016introduction}.}

The samples were dilutions of industrial car paint (WBC-713T No.202, Kansai Paint) with deionized water. A total of eight different test liquids were used, with the undiluted solution as 100 wt\% and seven solutions from 60--90 wt\% at 5 wt\% increments.
The fluids were stirred for 10--15 min using a magnetic stirrer.

\begin{table}[hbtp]
\caption{\RR{Surface tension of test liquids $\gamma$ and fitting parameters ($\eta_0$, $\alpha$, $\beta$) for various concentration of coating $\phi$.}}
\label{table:data_type}
\centering
\resizebox{\textwidth}{!}{
\begin{tabular}{S c S c c S c c}
\hline
$\phi$ [wt\%] & $\gamma$ [mN/m] & \RR{$\eta_{0,E}$ [mPa$\cdot$s]} & \RR{$\alpha_E$ [-]} & \RR{$\beta_E$ [-]} & \RR{$\eta_{0,S}$ [mPa$\cdot$s] }& \RR{$\alpha_S$ [-]} & \RR{$\beta_S$ [-]} \\
\hline
100 & 29.80 &8854 & -1.8$\times10^{-1}$ & 1.9$\times10^{-2}$ 
& 42940& 3.1$\times10^{1}$& 1.0$\times10^{-3}$\\
90 & 29.53 & 2726& -4.8$\times10^{-2}$ & 6.0$\times10^{-2}$ & 12110& 1.8$\times10^{1}$& 1.2$\times10^{-3}$\\
85 & 27.37 & 2094& -4.1$\times10^{-2}$ & 2.0$\times10^{-2}$ & 3711& 7.4$\times10^{0}$& 3.5$\times10^{-3}$\\
80 & 29.80 & 1372&3.3$\times10^{-2}$ & 3.7$\times10^{-2}$ & 1375& 3.9$\times10^{0}$& 9.0$\times10^{-3}$\\
75 & 27.35 & 620& 1.6$\times10^{-2}$ & 8.0$\times10^{-2}$ & 613& 2.7$\times10^{0}$& 1.6$\times10^{-2}$\\
70 & 28.24 & 1010& 	2.9$\times10^{0}$ & 1.3$\times10^{-2}$ & 137& -6.6$\times10^{-1}$& 3.9$\times10^{-2}$\\
65 & 27.88 &160& 2.3$\times10^{-3}$& 3.7$\times10^{-1}$ & 105& 3.3$\times10^{-1}$& 5.9$\times10^{-2}$\\
60 & 27.34 & 3976& 	1.6$\times10^{-1}$& 1.5$\times10^{-2}$ & 44& 2.7$\times10^{-1}$& 5.3$\times10^{-2}$\\
\hline
\label{Table:surface_tension}
\end{tabular}
}
\end{table}


\section{Results and discussion}
\label{S:4}
In this section, we present results for the filament radius (\S 4.1), extensional viscosity (\S 4.2), and shear viscosity (\S 4.3). Finally, we discuss the conditions in which the power-law relationships are applicable (\S 4.4).

\subsection{Filament radius}
\label{sec:radius_coating}
\begin{figure}[H]
\includegraphics[scale=0.54]{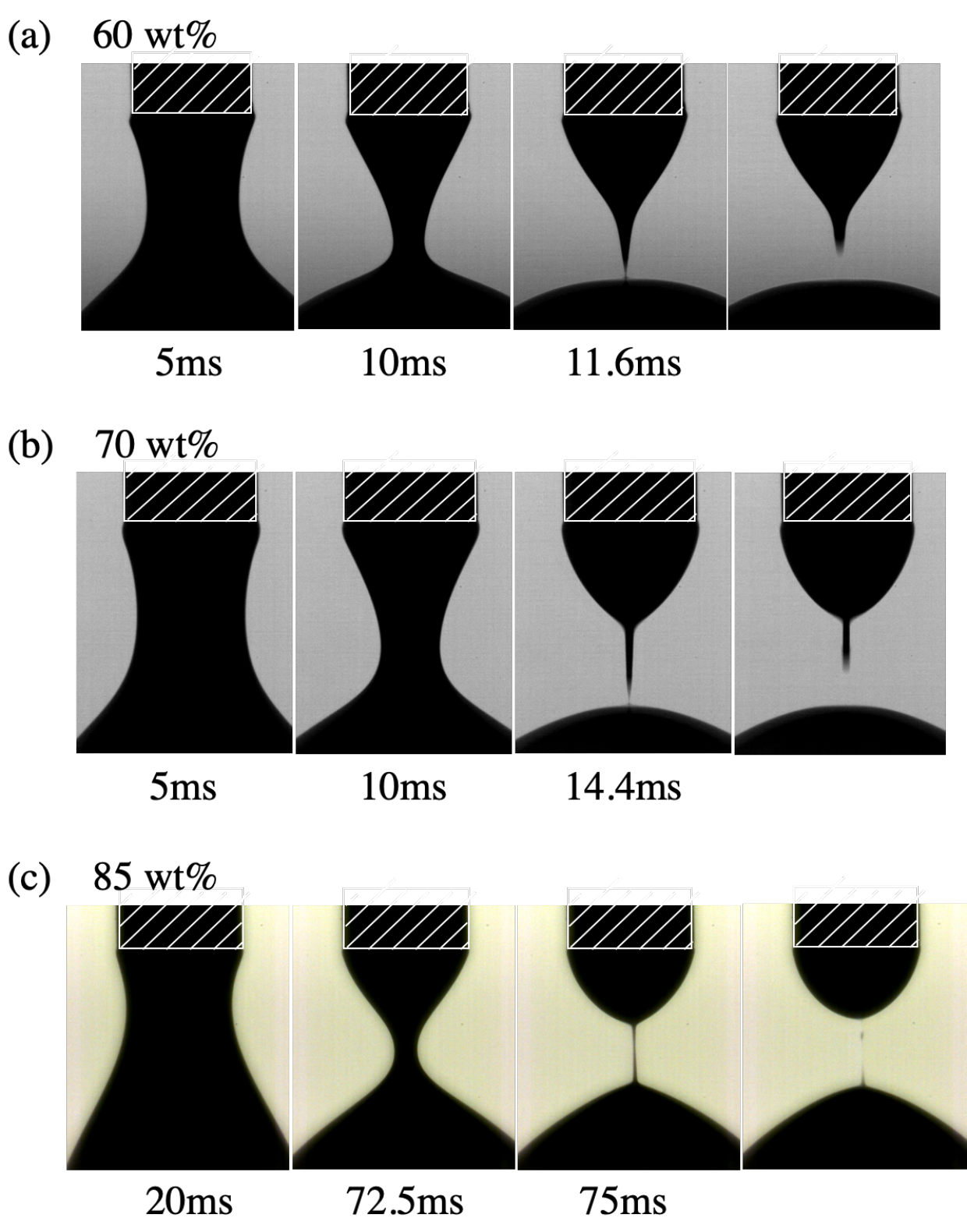}
\centering
\caption{Sequences of images showing liquid filament thinning in (a) 60 wt\% (low viscosity), (b) 70 wt\%, (c) 85 wt\% (high viscosity) test fluids. The third image is taken just before pinch-off and the final image is the next frame after the third image. The white shaded area is the nozzle.}
\label{Fig:coating_image}
\end{figure}

\begin{figure}[H]
\centering
\includegraphics[scale=0.45 ]{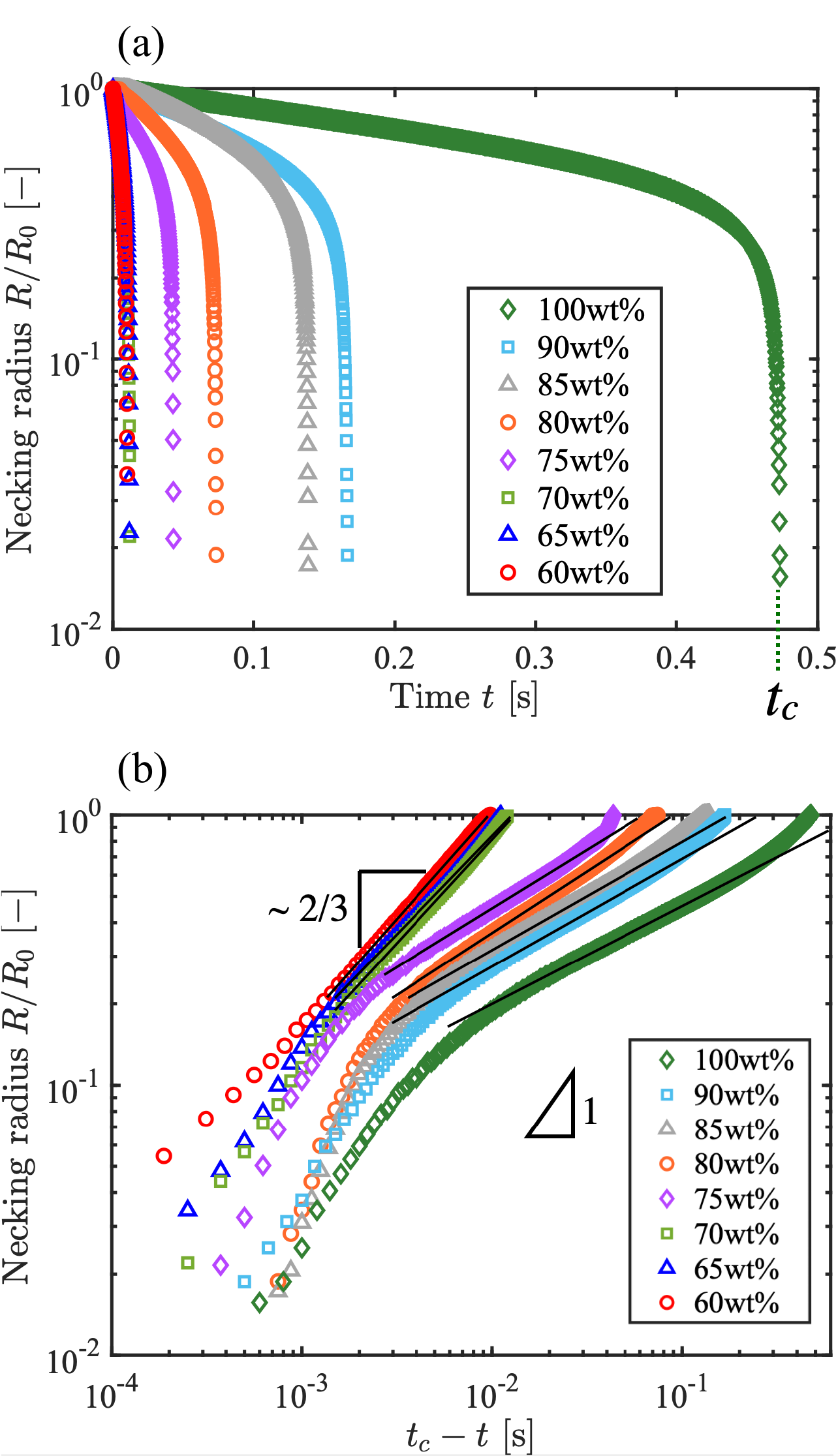}
\caption{(a) Time evolution of dimensionless filament radius $R/R_0$ of test fluids. $R$ is the smallest radius of the liquid filament. The time at which the dimensionless filament radius $R/R_0=1$ is set to $t=0$. The legend shows the concentration of test fluids. $t_c$ is the pinch-off time. (b) Dimensionless filament radius $R/R_0$ versus $t_c - t$. Solid lines show the results of the power-law approximation in the mid-necking stage.}
\label{Fig:coating_Rvstime}
\end{figure}

Image sequences of liquid filaments for various test fluids are shown in Fig.~\ref{Fig:coating_image}. 
Using the image sequences, the minimum filament radius $R$ was measured and the time evolution of the dimensionless radius $R/R_0$ of the liquid filament was obtained (Fig.~\ref{Fig:coating_Rvstime}(a)).
At higher concentrations, the time $t_c$ until pinch-off is longer.
Because $R$ for a power-law fluid can be expressed by Eq.~\eqref{eq11}, $R/R_0$ is plotted on a double-logarithmic graph with $t_c-t$ on the horizontal axis (Fig.~\ref{Fig:coating_Rvstime}(b)).
At lower concentrations ($<$70 wt\%), $R$ changes \R{according to a power law with an exponent of 2/3} throughout the entire necking process.
At higher concentrations ($>$ 75 wt\%), $R/R_0$ changes in three stages: early necking, mid-necking, and just before pinch-off. This trend, which has also been observed in a previous study~\cite{Clasen2012}, suggests that \RR{inertial acceleration is dominant} on the filament in the initial stage of necking, while viscous forces become dominant in the middle stage. 
In the case of suspensions, the viscosity decreases in the thinned filaments just before pinch-off, which accelerates the necking and results in faster breakage compared with Newtonian liquids of the same initial viscosity~\cite{Clasen2012}.
\R{McIlroy et al. \cite{mcilroy2014modelling} reported that a transition to the accelerated thinning regime starts at the point where the filament radius is approximately \RR{five times the size of the suspended particles}. This acceleration is attributed to the geometry of the filament. At this regime, the filament has high curvature. Therefore, thinning is accelerated in order to \RR{conserve volume} \cite{mcilroy2014modelling}. As the necking radius \RR{approaches} the particle size, the transition occurs from \RR{the} mid-term to \RR{the} final regime.} To discuss the \R{apparent} extensional viscosity in this experiment, as explained in Section 3, we focus on the mid-necking stage in which the viscous force is dominant. 
In this stage, we performed a power-law approximation (solid line in Fig.~\ref{Fig:coating_Rvstime}(b)) using Eq.~\eqref{Table:coating_exponent} and obtained the power-law exponent $n$ (Table \ref{Table:coating_exponent}).
The exponent for the time evolution of the filament radius increased from \R{concentration of }100 to 70\%. On the contrary, the value of $n$ tends to decrease \R{as the dilution increases} from \R{concentration of} 70 to 60\%, approaching the value of a Newtonian fluid ($n$ = 2/3) where inertial \RR{acceleration} dominates. This result suggests that the behavior of test fluids at 60--70\% is clearly different from that of power-law fluids.

\begin{table}[H]
\caption{Exponents of the test liquid. The filament radius exponent $n$ is obtained from a power-law approximation of the time evolution of the filament radius. The \R{apparent} extensional viscosity exponent $l$ and shear viscosity exponent $m$ are obtained by approximating the experimental results using Eqs.~\eqref{eq:extensionalvis_fit} and \eqref{eq:shearvis_fit}, respectively. \R{The exponents includes approximately error of 6.5\% using the way based on \S 3.} }
\centering
\begin{tabular}{Sccc}
\hline
{Concentration} & {Exponent of }& {Exponent of} & {Exponent of}
\\ {[wt\%]}
& {filament radius}
& {\R{apparent} extensional viscosity}
& {shear viscosity}
\\ 
& {$n$ [-]}
& {$l$ [-]}
& {$m$ [-]} 
\\
\hline
100 & 0.365 & \RR{$-0.768$} & \RR{$-0.736$} \\
90 & 0.421 & \RR{$-0.829$} & \RR{$-0.681$} \\
85 & 0.415 & \RR{$-0.721$} & \RR{$-0.635$} \\
80 & 0.474 & \RR{$-0.712$} & \RR{$-0.591$} \\
75 & 0.514 & \RR{$-0.776$} & \RR{$-0.537$}\\
70 & 0.755 & \RR{$-0.321$} & \RR{$-0.393$}\\
65 & 0.733 & \RR{$-1.304$} & \RR{$-0.456$}\\
60 & 0.721 & \RR{$-0.999$} & \RR{$-0.271$}\\
\hline
\label{Table:coating_exponent}
\end{tabular}
\end{table}

\subsection{Extensional viscosity}
\label{sec:Extensional viscosity}

\begin{figure}[H]
\centering
\includegraphics[scale=0.6]{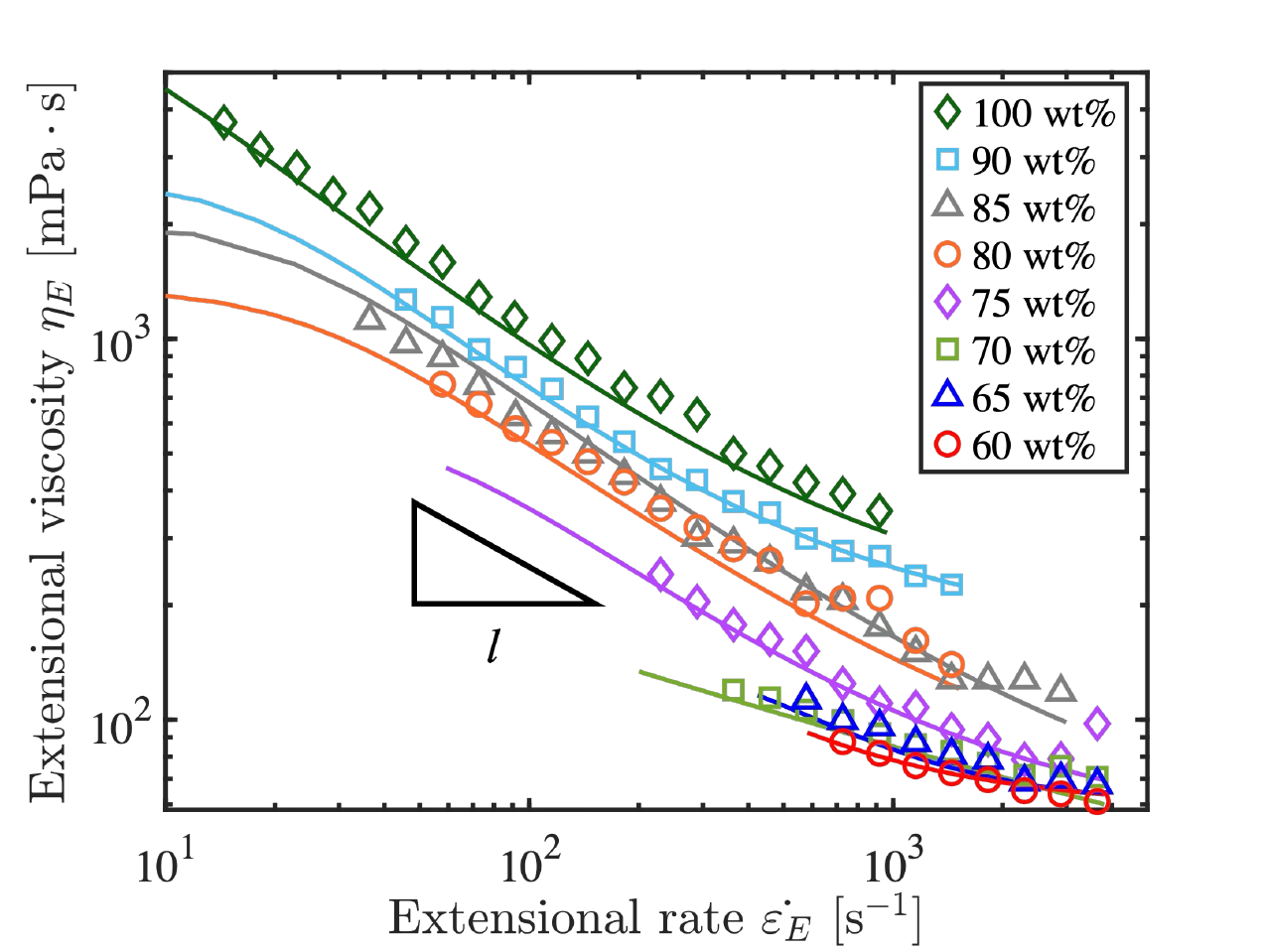}
\caption{\R{Apparent} extensional viscosity versus extensional rate of diluted coating measured using CaBER-DoS. The legend shows the concentrations. Solid lines show the results of \RR{fitting to} Eq.~\eqref{eq:extensionalvis_fit}. This approximation yields the power-law exponent $l$.}
\label{Fig:coating_extensional}
\end{figure}

The results for the \R{apparent} extensional viscosity as a function of the extensional rate of the test fluid are shown in Fig.~\ref{Fig:coating_extensional}.
The test fluids exhibit strain-thinning for $\dot{\varepsilon} < 1000\ \rm{s}^{-1}$.
This is similar to the results of O'Brien and Mackay~\cite{O'brien2002}, who experimentally showed that the extensional viscosity of a suspension undergoes \RR{strain rate} thinning at extensional rates of 10--1000 $\rm{s}^{-1}$. 
In our experiments, higher concentrations exhibit higher \R{apparent} extensional viscosity, but low-concentration solutions (60--70 wt\%) have a similar \R{apparent} extensional viscosity at every extensional rate.
The obtained \R{apparent} extensional viscosity was fitted to Eq.~\eqref{eq:extensionalvis_fit} to obtain the power-law exponent $l$ (Table \ref{Table:coating_exponent}).
At high concentrations ($<$80 wt\%), the absolute value of $l$ tends to \R{decrease with dilution}.
Below 70 wt\%, the power-law exponent $l$ increases with \R{increasing with dilution}.

We now discuss why the \R{apparent} extensional viscosity and power-law exponent $l$ exhibit different trends in high- and low-concentration solutions.
In deriving the \R{apparent} extensional viscosity, we assumed a liquid filament in which the viscous force is dominant and the inertial \RR{acceleration} is negligible.
However, in fact, inertial \RR{acceleration} are not negligible in low-viscosity solutions as discussed before.
Therefore, at low concentrations (60--70 wt\%), there may be a larger difference in the apparent extensional viscosity than in the true extensional viscosity.
The following discussion will proceed with the data for 80\% or more of the data being relied upon and 60--70\% being taken as reference.

\subsection{Shear viscosity}
\begin{figure}[H]
\centering

\includegraphics[scale=0.6]{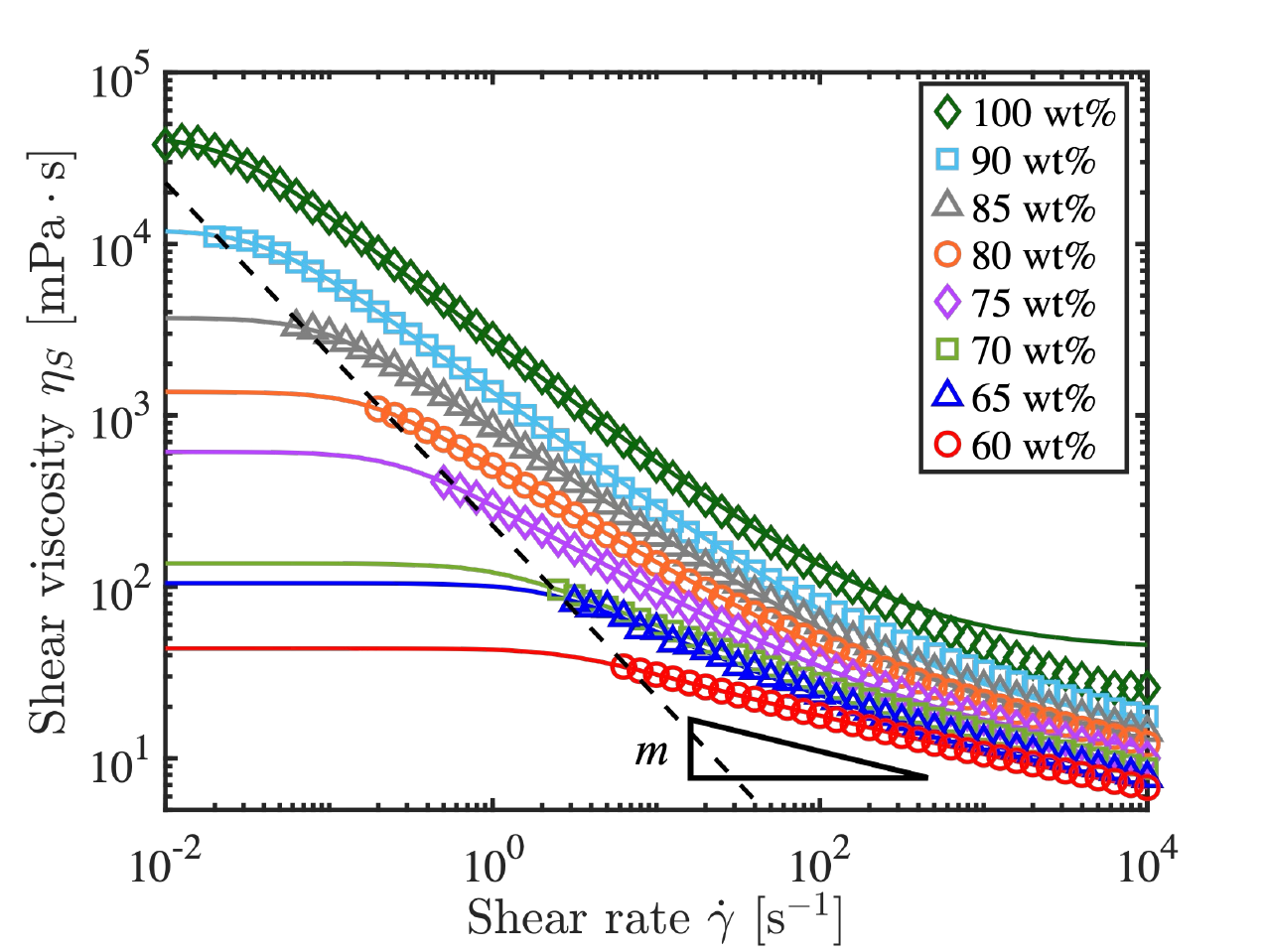}
\caption{Shear viscosity versus shear rate of test fluids measured using shear rheometer. The legend shows the concentration of the coating. Solid lines show the \RR{fitting to} Eq.~(\ref{eq:shearvis_fit}). The approximations yield the power-law exponent $m$. The dashed line shows the low-torque limit, Eq.~\eqref{eq:23}, and we discuss the region to the right of the dashed line.}
\label{Fig:coating_shear}
\end{figure}

The shear viscosities of the test fluids, as measured with a shear rheometer, are shown in Fig.~\ref{Fig:coating_shear}.
The reliable range according to Eq.~\eqref{eq:23}, which is the region to the right of the dashed line in Fig. 7, is now discussed. 

In a previous study on rheological measurements of \A{coating,} the shear viscosity was found to exhibit a shear-thinning property~\cite{Ascanio2006}. 
In our experiments, the shear viscosity was measured using a shear rheometer, and it was found that the test fluids exhibited shear-thinning, similar to the results of Ascanio~\cite{Ascanio2006}.

The shear viscosity decreases \R{on increasing dilution} of the test fluids. 
Thus, the shear viscosity increases monotonically as the viscous force increases, i.e., with \R{on decreasing dilution}.
The power-law exponent $m$ was obtained by fitting Eq.~\eqref{eq:shearvis_fit} to the obtained shear viscosities (Table \ref{Table:coating_exponent}).
\R{Lower dilutions} of the test fluid have larger absolute values of the power exponent $m$. Thus, at \R{lower dilutions}, the shear-thinning property is stronger.
Conversely, $m$ tends to zero as \RR{dilution of the test fluid increases}, and the shear viscosity approaches a constant value, similar to a Newtonian fluid.

\subsection{Relationship between extensional and shear rheology: power-law exponents}
\begin{figure}[H]
\includegraphics[scale=0.6]{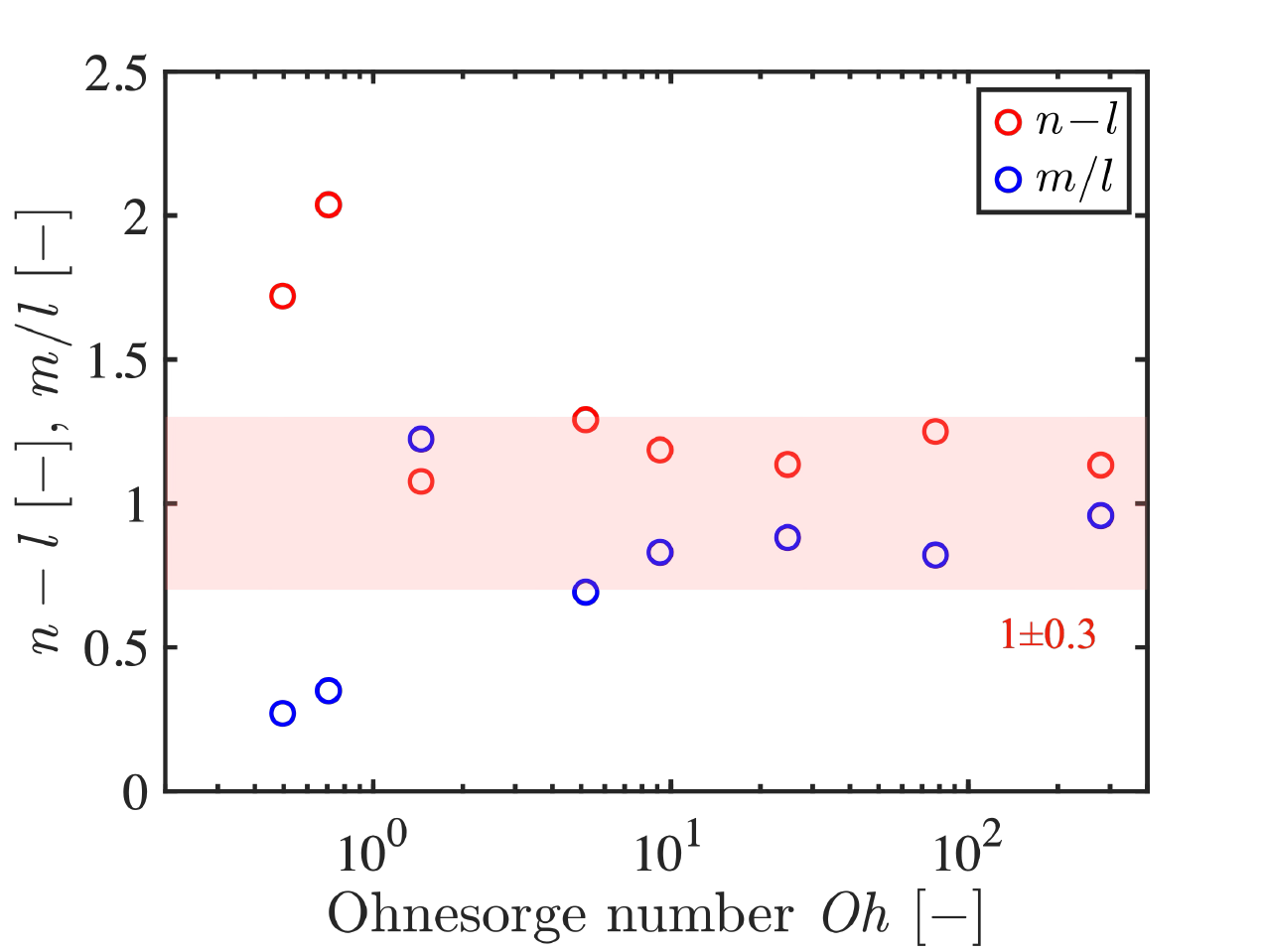}
\centering
\caption{$n-l$ (red) and $m/l$ (blue) versus Ohnesorge number $Oh$. The area highlighted in red indicates the range of $1\pm 0.13$ for $n-l$ and $m/l$.}
\label{Fig:n_l_m}
\end{figure}

From our experiments, the power-law exponents $n$, $l$, and $m$ for the filament radius, \R{apparent} extensional viscosity, and shear viscosity were obtained. 
In the following, we explain the $Oh$ conditions under which Eqs.~\RR{\eqref{eq:l} and \eqref{eq:relationship_m_l}} hold. 
Figure \ref{Fig:n_l_m} shows the values of $n-l$ and $m/l$.
It is clear that $n - l = 1$ for $Oh > 1$ in the range of $\pm$\RR{29}\%.
Our experimental results indicate that the relation between the power-law exponents of the \R{apparent} extensional viscosity and the radius of the liquid filament for $Oh>1$ is valid, as shown numerically by Suryo \R{and Basaran}~\cite{Suryo2006}.
For $Oh > 1$, the power-law exponents for the \R{apparent} extensional viscosity and shear viscosity obey the relation $m / l = 1$ in the range of $\pm$\RR{31}\%, which is in agreement with the power-law expression based on the Carreau model.
Below 65 wt\%, $n-l$ becomes larger than 1 and $m/l < 1$.

From the above results, it is clear that $Oh > 1$ is a condition for the applicability of the viscous power-law expression with respect to the liquid filament radius, apparent extensional viscosity, and shear viscosity.
This result may be useful in predicting the \R{apparent} extensional viscosity from shear viscosity measurements when the relationship between the shear and extensional viscosities follows the Carreau model in the viscosity-dominated region.
As shear viscosity measurements are easier than extensional viscosity measurements, any discussion of the extensional viscosity will be easier if we can predict the \R{apparent} extensional viscosity from measured values of the shear viscosity.
For example, because the liquid jet pinches off after the extension of the filament, the results of this study will be useful for controlling the injection volume of the jet, which has practical applications in on-demand printing technologies.

\section{Conclusion}
\label{S:5}
This study has experimentally investigated the relationship between the shear and extensional rheology of power-law fluids.
A previous study based on numerical simulations by Suryo and Basaran~\cite{Suryo2006} suggested that the power-law relationship of extensional rheology is applicable when $Oh > 1$, but this condition was not experimentally investigated. 
In addition, this previous study only simulated the necking behavior, and did not discuss the shear viscosity.
We experimentally investigated the relationship between the power-law exponents relating to the shear and extensional rheology. 
For the extensional rheology measurements, we used CaBER-DoS, which is capable of measuring fluids with lower viscosity than other measurement methods.
The shear rheology was measured using a shear rheometer.
The applicable conditions were investigated by changing the \R{dilutions} of the power-law fluids.
The measured \R{apparent} extensional and shear viscosities were compared with the theoretical relationships given by the Carreau model.

Our experimental results revealed that the power-law expression for the filament radius, apparent extensional viscosity, and shear viscosity is applicable in the case of at least $Oh > 1$. In this condition, the exponents for shear and extensional viscosity become same. \D{In addition, the difference between exponents of filament radius and that of \R{apparent} extensional viscosity become constant value of 1 for $Oh > 1$.} 
The results presented in this paper will be useful in predicting the \R{apparent} extensional viscosity from the shear viscosity, and will contribute to industrial applications such as on-demand printing. 

\R{Our results provide experimental evidence for the criteria of $Oh$, which were also suggested numerically by Suryo and Basaran~\cite{Suryo2006}. However, this study is limited to a single fluid, industrial car paint, modeled as a power-law fluid. Therefore, future work should examine other fluids, such as aqueous solutions of xanthan gum.}

\section*{ACKNOWLEDGMENTS} 
This work was supported by JSPS KAKENHI (Grant Nos. JP20H00222 and JP20H00223.), JST SBIR program (Grant No. JPMJST2355) and ASTEP program(Grant No. JPMJTR212C). We thank Dr. Jingzu Yee for English editing and proof reading, Dr. Prasad Sonar and Dr. Pradipto for proof reading.

\bibliography{Ref}
\bibliographystyle{elsarticle-num-names}

\end{document}